\documentclass[conference]{IEEEtran}
\IEEEoverridecommandlockouts
\usepackage[hyphens]{url}
\usepackage{listings}
\usepackage{cite}
\usepackage{hyperref}
\usepackage{amsmath,amssymb,amsfonts}
\usepackage{algorithmic}
\usepackage{graphicx}
\usepackage{textcomp}
\usepackage{xcolor}
\def\BibTeX{{\rm B\kern-.05em{\sc i\kern-.025em b}\kern-.08em
    T\kern-.1667em\lower.7ex\hbox{E}\kern-.125emX}}

\makeatletter
\def\lst@makecaption{%
  \def\@captype{table}%
  \@makecaption
}
\makeatother

\begin{document}

\title{An In-memory Embedding of CPython for Offensive Use}

\author{\IEEEauthorblockN{Ateeq Sharfuddin, Brian Chapman, Chris Balles}
\IEEEauthorblockA{
\textit{SCYTHE}\\
\{ateeq, brian, chris\}@scythe.io 
}
}

\maketitle

\begin{abstract}
We offer an embedding of CPython that runs entirely in memory without ``touching'' the disk. This in-memory embedding can load Python scripts directly from memory instead these scripts having to be loaded from files on disk. Malware that resides only in memory is harder to detect or mitigate against. We intend for our work to be used by security researchers to rapidly develop and deploy offensive techniques that is difficult for security products to analyze given these instructions are in bytecode and only translated to machine-code by the interpreter immediately prior to execution. Our work helps security researchers and enterprise Red Teams who play offense. Red Teams want to rapidly prototype malware for their periodic campaigns and do not want their malware to be detected by the Incident Response (IR) teams prior to accomplishing objectives. Red Teams also have difficulty running malware in production from files on disk as modern enterprise security products emulate, inspect, or quarantine such executables given these files have no reputation. Our work also helps enterprise Hunt and IR teams by making them aware of the viability of this type of attack. Our approach has been in use in production for over a year and meets our customers' needs to quickly emulate threat-actors' tasks, techniques, and procedures (TTPs).

\end{abstract}


\section{Introduction}
Different classes of threat-actors have different motivations, and therefore, produce and utilize different types of malware to complete their goals. Nation-states are the most sophisticated threat-actors and go through great lengths to ensure the end-user is not alerted to their intrusions. As a consequence, whereas cyber-criminals are not so concerned about leaving artifacts of compromise behind, malware produced by nation-states aim to leave no artifact behind and have built-in security hindering forensic analysis. One method threat-actors use is to keep much of the running malware only in volatile memory, for example, a system's RAM. This way, when the system shuts down or is restarted, there is no disk artifact of the intrusion as state is lost in volatile memory when power is removed from the system.

Python is a popular interpreted high-level programming language~\cite {zdnet2020} with an emphasis on readability and an abundance of reusable packages produced by the community. CPython is the default and most widely used implementation of the Python language, written in C and Python. With the {\small{\texttt{ctypes}}} package, CPython is as expressive as C, allowing non-interpreted native instruction execution. One common scheme unsophisticated threat-actors use is to write their malware in Python and package the malware and all its dependencies into an executable using packagers like Py2Exe~\cite{py2exe} or PyInstaller~\cite{pyinstaller}. These malware are easy to construct and are executed when a coerced end-user double-clicks and runs the packaged executable file from disk. Security products have collected and analyzed numerous sample executable files of this scheme. As a result, security products have produced signatures for this scheme and are able to successfully block all malware using this scheme.

In this paper, we contribute the first embedding of CPython interpreter that runs entirely in memory without ``touching'' the disk. We also contribute a scheme to support loading and running Python modules including native C extensions, also residing only in memory, onto this interpreter. We are able to achieve these results without needing any file or dependency to reside on disk, unlike the Py2Exe or PyInstaller packaging schemes. We have not come across any other in-memory embeddings of Python. Constructing this CPython embedding was an engineering effort: We claim no new algorithms or techniques. Our paper details the necessary steps to successfully load and utilize an in-memory embedding of CPython. 

Our goal is to assist security researchers and enterprise Red Teams. Enterprise Read Teams are responsible for emulating numerous classes of threat-actors. Red Teams prefer to repurpose existing malware or open-sourced security experiments. Red Teams have deadlines to meet and run campaigns with periodicity: As such, they want to emulate different threat-actors quickly, and this is more time-consuming to accomplish in a binary compiled from C or assembly than executing statements in a Python interpreter. These malware also need to evade Hunt and Incident Response teams until the objectives for the campaign are at least partially met. Red Teams also have difficulty running malware from files on disk as modern enterprise security products emulate, inspect, or quarantine such executables given that these files have no reputation. A month-long campaign effort can be rendered worthless on the first day of use if a Red Team's artifact is immediately caught upon execution by the Hunt or IR teams. We offer this information in the interest of raising awareness of the viability of this type of attack. As a simplification, we do not cover any additional security or obfuscation mechanisms that a threat-actor would use, for example, changing bytecode mappings or preventing memory dumps, nor do we cover how to land onto the virtual address space of a process: These well-known topics have been covered by other researchers ~\cite{blackberry2019pyxie, pearce2018covert}. Both Windows and POSIX implementations are in use in production. As a simplification, we only extensively cover the steps for Windows. However, similar steps apply to POSIX, and we describe differences in Section~\ref{sec:posix}.

Section~\ref{sec:background} describes prior research relevant to this paper. Section~\ref{sec:embedding} details the in-memory CPython interpreter. Section~\ref{sec:initializing} details loading Python scripts from memory into this in-memory interpreter. Section~\ref{sec:c_extensions} details in-memory dynamic loading of native Python C Extensions into this in-memory interpreter. Section~\ref{sec:evaluation} covers emulation of a few security techniques with this interpreter. Section~\ref{sec:future} covers future work.

\section{Background}
\label{sec:background}
We cover prior work relevant to this paper in this section. Specifically, we cover related Python embeddings, dynamic loading techniques, and in-memory loading of Python modules. We utilize the in-memory dynamic loading technique derived from a prior work~\cite{skape2004remote} in order to load native C Python Extensions into memory without ``touching'' the disk: This does not need to be reinvented.

\subsection{Contemporary Python Embeddings}
Numerous malware have been publicized that use the Py2Exe or PyInstaller methods~\cite{cyborgsecurity}. Signatures have been produced for these malware, and often security products simply quarantine any Py2Exe or PyInstaller-generated executable without inspecting if the encapsulated payload is malicious or benign. One malware was publicized in~\cite{blackberry2019pyxie} where the malware-authors recompiled the CPython interpreter and remapped the opcodes to make analysis difficult: In this case, this recompiled interpreter was still an executable file on disk. In~\cite{kholia2013looking}, the authors take a peek inside Dropbox's executable file and use a reflective loader~\cite{stephenfewer} to inject their shared library into the Dropbox process's address space: The Dropbox executable file was also a Python interpreter. These listed cases differ from our approach: We show how to download the interpreter into a location in memory and then run the interpreter from there. Our approach does not ``touch'' the disk, meaning no intermediate step exists where the interpreter or any dependency is saved as a file to disk in order for the modified CPython interpreter to function. 

\subsection{Dynamic Loading}
The documented way to load a shared library at runtime (also known as dynamic loading~\cite{lu1995elf}) is to use functions provided by operating system APIs such as {\small{\texttt{LoadLibrary}}} on Windows or {\small{\texttt{dlopen}}} on Linux. These functions require the shared library to be a file on disk. However, it is entirely possible to load these directly from a location in memory, and well-known approaches are described in~\cite{skape2004remote, levine1999linkers}. Implementations are available online~\cite{stephenfewer, fancycode}. We employ one of these approaches to load the Python C extensions, which are native shared libraries, directly from a process's address space. The interpreter is compiled only once, and you can load modules on-demand into the interpreter at run-time with this approach. An alternative approach would have been to ``freeze''~\cite{freeze} these Python C extensions and compile them into the interpreter. However, this is not a flexible approach: You must recompile the interpreter each time for a different set of ``frozen'' modules.

\subsection{Python module Loaders}
The official stock CPython interpreter loads modules and packages from files on disk using its \textit{PathFinder} importer object. The interpreter also contains a frozen \textit{zipimport}~\cite{zipimport} module that is limited to importing a single package or module from a ZIP file on disk. This \textit{zipimport} cannot import Python C extensions, either. We constructed our {\small \texttt{cba\_zipimport}} derived from \textit{zipimport} that can load multiple packages, modules, and Python C extensions from an in-memory ZIP archive.

\section{In-memory Embedding of Python}
\label{sec:embedding}
We cover the concept of embedding the Python interpreter in another program in this section. The Python interpreter is available as a shared library. We cover loading a shared library from memory, and although this is prior art, we feel it is nonetheless important to summarize. Note that using an in-memory dynamic loading method~\cite{skape2004remote} to load this Python interpreter shared library into memory is not enough to initialize and start invoking Python scripts. We list the additional necessary steps that lead to a successful initialization of the interpreter. We concluded that these were the only steps necessary in an iterative fashion by running the CPython interpreter through a debugger observing failures, reading the CPython source code, and reading official CPython documentation, in that order.

\subsection{Embedding Python}
The Python interpreter was designed to be embedded into applications. Embedding allows an application developer to implement some application functionality in the Python programming language rather than in C or C++~\cite{embedpython}. An application initializes the embedded Python interpreter with a call to {\small \texttt{Py\_Initialize}}. The interpreter can then be called from any part of the application. Simple and pure embeddings are covered in detail in~\cite{embedpython}. The Python interpreter itself is implemented in a shared library. For example, the Python 3.8 core shared library on Windows is python38.dll, and the main Python application (python.exe) loads this shared library and provides a read-eval-print loop (REPL) interface to the user. To embed the Python interpreter into an application, simply link this Python core shared library during compile-time (using static linking or dynamic linking) or load it from the application at run-time (with dynamic loading).

\subsection{Loading Shared Libraries from Disk}
There are three mechanisms a program can choose to use other software~\cite{tldp}: static linking, dynamic linking, and dynamic loading. Dynamic loading allows a program to, at run time, load a shared library into memory, retrieve addresses of functions and variables, execute these functions or access these variables, and unload the library from memory when the application no longer needs the library. Shared libraries such as Dynamic-Link Libraries (.dll files) on Windows or Shared Object Libraries (.so files) on Linux can be dynamically loaded at runtime. As a simplification, Windows API provides a C function {\small \texttt{LoadLibrary(FilePath)}} to load a shared library into a process's address space. The C function {\small \texttt{GetProcAddress(ModuleHandle, FunctionName)}} can then be used to retrieve the address of an exported function (or variable) in the shared library. If the function is found, the application code can now call this function. Finally, the application can call {\small \texttt{FreeLibrary(ModuleHandle)}} to unload the shared library from the process's address space when no longer in use~\cite{msdnmag}. In the POSIX standard API, the equivalent C functions are {\small \texttt{dlopen}}, {\small \texttt{dlsym}}, and {\small \texttt{dlclose}}~\cite{tldp}, respectively.

\subsection{Loading Shared Libraries from Memory}
Although official documentation exposes APIs where the shared libraries must be files on disk, research has been published~\cite{skape2004remote} that demonstrate that shared libraries can in fact be dynamically loaded directly from memory and reference implementations exist~\cite{fancycode}. These research provide details necessary to reimplement the operating system's loader and offer three replacement C functions: {\small \texttt{LoadLibraryFromMemory(ModuleBytes)}}, {\small \texttt{GetMemoryProcAddress(Handle, FunctionName)}}, and {\small \texttt{FreeInMemoryLibrary(Handle)}}. With such a reimplementation of the operating system's loader we can load the Python core shared library directly into a process's address space.

\subsection{Preparation}
Prior to loading the shared library into a process's address space, the process itself must be compromised: We write and execute instructions to download the in-memory loader code and the shared library, and proceed to loading this shared library. We do not detail how to compromise a process in this paper: An extensive amount of prior art exists in this area~\cite{bh2017, bh2019} and compromising a process does not have a one-solution-fits-all answer nor is this the topic of this paper. This paper assumes you already control the instruction pointer in a process, wrote to memory to download the loader code and the shared library from the network, and have successfully loaded this shared library with a call to {\small \texttt{LoadLibraryFromMemory}}. We describe modifications needed to specifically run the Python interpreter after loading its shared library.

\section{Initializing the Interpreter}
\label{sec:initializing}
After successfully loading the Python core shared library into memory, we now need to initialize the embedded interpreter in it. Instead of using the traditional {\small \texttt{Py\_Initialize}} call to initialize the embedded interpreter we use {\small \texttt{Py\_InitializeFromConfig}}. The configuration structure is initialized with a call to {\small \texttt{PyConfig\_InitIsolatedConfig}}. In isolated mode the embedded Python interpreter ignores environment variables, global configuration variables, command line arguments, and the user site directory. This is of particular importance to us: We want to run a Python interpreter that does not check for an existing Python installation on the system. Specifically, not all systems will have Python pre-installed, the version of Python pre-installed on the system may not match the version of the embedded interpreter, or loading packages from the pre-installed Python will produce auditable events (for example, Sysmon Event 11 FileCreate ~\cite{sysmon11}) that may raise alarms. As an example, if hypothetically, the system calculator (calc.exe) starts an embedded Python, and it reads a collection of py files in rapid succession, it is suspicious. 

\subsection{Required Modules}
Starting the embedded Python interpreter in-memory in isolated mode will fail, as the interpreter needs to load modules and packages such as \textit{encodings}, \textit{codecs}, \textit{abc}, etc. and cannot find these in the path: We must make changes to the embedded Python interpreter so that it can load modules and packages it needs directly from memory instead. We put all the requisite packages and modules in a ZIP archive and transmit this ZIP archive along with the interpreter to the process (as explained in Section~\ref{bundling}). The embedded interpreter loads the necessary packages and modules from this in-memory ZIP archive. With this approach, we do not have to transmit each package or module individually. The CPython provided builtin \textit{zipimport}~\cite{zipimport} module can import Python modules and packages from ZIP files. However, \textit{zipimport} does not support importing from ZIP archives in memory.  Nor does \textit{zipimport} support importing multiple modules and packages residing in the same ZIP archive or importing Python C extensions in a ZIP archive. We wrote a derivative of Python's builtin \textit{zipimport} module, to remove these limitations: The \texttt{cba\_zipimport} module loads packages and modules from ZIP archives in memory and can also load Python C extensions.

\subsection{cba\_zipimport MetaPathFinder}
PEP 302~\cite{pep302} describes the concept of \textit{metapath} and its implementation provides a list of importer objects (in \textit{sys.meta\_path}) that are traversed to import packages and modules. The last object in \textit{sys.meta\_path} is the \textit{PathFinder} object, which loads modules and packages from files on disk. The entire import system is exposed via \textit{sys.meta\_path} with no implicit machinery~\cite{pythonimport}. We implemented the importer {\small \texttt{cba\_zipimport}}, and inserted it into \textit{sys.meta\_path}. This way, when the interpreter traverses \textit{sys.meta\_path} to load a module or package, it will inquire {\small \texttt{cba\_zipimport}} to load this module. The order in which {\small \texttt{cba\_zipimport}} will be called depends on its position in the \textit{sys.meta\_path} list.

The {\small \texttt{cba\_zipimport}} module provides an implementation of \textit{importlib.abc.MetaPathFinder} as \emph{finder} and an implementation of \textit{importlib.abc.Loader} as \emph{loader}. The interpreter will use this \emph{finder} to locate the \emph{loader} for a module. The interpreter uses the \emph{loader} to then load the module. The {\small \texttt{cba\_zipimport}} module also provides functionality, which when invoked inserts {\small \texttt{cba\_zipimport}} as an importer object into \textit{sys.meta\_path}. The underlying implementation of parsing ZIP archives in {\small \texttt{cba\_zipimport}} remains exactly the same as Python's \textit{zipimport}.

\subsection{Adding to Builtin Frozen modules}
We ``freeze''~\cite{freeze} {\small \texttt{cba\_zipimport}} and add {\small \texttt{cba\_zipimport}} to the table referenced by {\small \texttt{PyImport\_FrozenModules}}. This way, the Python interpreter already will have the ``frozen'' {\small \texttt{cba\_zipimport}} and will not try to search for it using \textit{PathFinder} on the system's disk.

\subsection {PyLifecycle Updates}
We slightly update pylifecycle.c: Prior to {\small \texttt{init\_importlib\_external}} calling {\small \texttt{\_PyImportZip\_Init}}, we add a call to {\small \texttt{\_PyCBAImport\_Init}} function. {\small \texttt{\_PyCBAImport\_Init}} imports our {\small \texttt{cba\_zipimport}} module using the interpreter's C function {\small \texttt{PyImport\_ImportModule}}. If import is successful, {\small \texttt{\_CBAZipImport\_Init}} calls the module's function {\small \texttt{install\_cba\_metafinder}}. This function instantiates a global instance of \emph{loader} and \emph{finder}, and inserts \emph{finder} in \textit{sys.meta\_path} at offset 2. That is, this \emph{finder} is the third in the list of finders to be called should the interpreter need to load a module: The two importer objects preceding this \emph{finder} are Python's {\small \texttt{BuiltinImporter}} and {\small \texttt{FrozenImporter}}. As an example, if the interpreter were looking for \textit{codecs.py}, it would first attempt to find its loader with \textit{BuiltinImporter}. If \textit{BuiltinImporter} cannot provide one, it would attempt to find one with \textit{FrozenImporter}. If \textit{FrozenImporter} cannot provide one, it would attempt to find one with \emph{finder}. If \emph{finder} could not provide one, the interpreter would attempt to find \textit{codecs.py} with \textit{PathFinder}.

\subsection{Bundling Python Libraries}\label{bundling}
Since a stock installation of Python contains a prepackaged collection of modules, we want the in-memory Python interpreter to also offer this same collection of modules. We create a ZIP file, cba\_python38\_lib.zip. We put everything under Python's Lib directory into cba\_python38\_lib.zip: These are standard Python modules, which have .py or .pyc extensions. We constructed an xxd.py, which behaves similar to the xxd application, and used this xxd.py to output a C source file with the contents of the aforementioned ZIP file stored in a C array. We compile this C source file into the Python core shared library.  When {\small \texttt{\_CBAZipImport\_Init}} calls the {\small \texttt{cba\_zipimport}} module's function {\small \texttt{install\_cba\_metafinder}}, this C array is provided as an input parameter.

\subsection{Result}
At this point, the in-memory Python interpreter has everything it requires to be loaded directly from memory. To summarize, the in-memory loader loads the Python core shared library hosting the Python Interpreter. We initialize the interpreter calling it with an isolated configuration. The interpreter initializes and loads modules, and initializes builtin modules and frozen modules. The {\small \texttt{cba\_zipimport}} module, which is frozen, is loaded and it's \emph{finder} is added to Python's \textit{sys.meta\_path} at offset 2. At this point, the interpreter requests to load modules, and this \emph{finder} resolves these requests and successfully loads them from the in-memory ZIP archive that is bundled into the interpreter. All the necessary modules are found, and the in-memory Python interpreter is now ready to execute Python statements.

\section{Python C Extensions}
\label{sec:c_extensions}
In specific scenarios, calling native code from Python is necessary. Operating systems come with a large number of APIs that aren't present in the Python interpreter, an example scenario being APIs to access hardware or operating system management functions. Similarly, there may be a reason where the Python interpreter needs to communicate with components written in another language that produces a C-style native application binary interface (ABI), an example scenario being some Java code exposed via Java Native Interface (JNI). Python C Extensions offer a way to extend Python itself: Implement native code in the extension and call it from the Python interpreter. For example, the \textit{ssl} Python package imports \textit{\_ssl}, which is a C extension linked with the native OpenSSL shared library that calls operating system APIs.

For our approach, when a Python C Extension depends on a shared library, we statically link the shared library into the extension. This rule excludes shared libraries that are guaranteed to exist on the system, for example, kernel32.dll on Windows. Using the previous example, the C extension \_ssl.pyd for Python 3.8 expects the shared library libcrypto-1\_1.dll to be in the path. In our approach, the in-memory interpreter cannot expect to resolve libcrypt-1\_1.dll from disk. Therefore, it is appropriate to statically link OpenSSL into \_ssl.pyd.

All the Python C Extensions that come bundled with the standard CPython interpreter are recompiled such that the non-system shared library dependencies are statically-linked. These C Extensions are put in a ZIP file cba\_python38\_pyd\_lib.zip. We use xxd.py to output a C array with the contents of the aforementioned ZIP file.  We then compile this C source file into the Python core shared library. {\small \texttt{\_PyCBAImport\_Init}} calls the {\small \texttt{cba\_zipimport}} module's function {\small \texttt{install\_cba\_metafinder}} with this C array as input parameter, similar to what we did for the Python modules that were not C extensions. With this step the in-memory interpreter has the same collection of modules and packages as a stock installation of the official Python interpreter.

\subsection{Loading C extensions in ZIP archives}
\subsubsection{The \_zip\_searchorder structure}
We update the {\small \texttt{\_zip\_searchorder}} structure in {\small \texttt{cba\_zipimport}} to track if a Python module is native. The {\small \texttt{cba\_zipimport}} is derived from Python's \emph{zipimport}: We describe the alterations we made in our derivation. The elements in each tuple are: suffix, is bytecode, is package, is native. In the structure as shown below, files with a .pyd extension are neither bytecode, nor a package, but are native.

\lstset{language=Python}
\lstset{frame=lines}
\lstset{caption={\_zip\_searchorder structure}}
\lstset{label={lst:code_direct}}
\lstset{basicstyle=\footnotesize}
\begin{lstlisting}
# (extension, isbytecode, ispackage, isnative)
_zip_searchorder = (
  (path_sep + '__init__.pyc', True, True, False),
  (path_sep + '__init__.py', False, True, False),
  ('.pyc', True, False, False),
  ('.py', False, False, False),
  ('.pyd', False, False, True),)
\end{lstlisting}

The functions {\small \texttt{\_get\_module\_info}} and {\small \texttt{\_get\_module\_code}} use this \textit{\_zip\_searchorder} structure to determine if the lookup resolves to a package, a module, or neither. The function {\small \texttt{load\_module}} calls {\small \texttt{\_get\_module\_code}} and has been updated to call the native loader if {\small \texttt{\_get\_module\_code}} indicates the filename in the ZIP archive being loaded is native.

\lstset{language=Python}
\lstset{frame=lines}
\lstset{caption={load\_module snippet}}
\lstset{label={lst:code_cba}}
\lstset{basicstyle=\footnotesize}
\begin{lstlisting}
  ...
  code, ispackage, modpath, isnative =
        _get_module_code(self, fullname)
  ...
  try:
    ...
    # existing path
    if not isnative:
        exec(code, mod.__dict__)
    else:
        # code is a native C extension
        mod = _native_code(fullname, code)
  ...
  return mod

\end{lstlisting}


The \textit{\_native\_code} function instantiates a \textit{ModuleSpec} for the C extension, and proxies the remaining work of loading the C extension to be performed by the builtin importer. The builtin import module was extended with an additional method {\small \texttt{create\_dynamic\_inmemory}}, which accepts a ModuleSpec instance for the extension to be loaded and a byte array containing the uncompressed bytes storing the extension. Functionally, the only difference between the new {\small \texttt{create\_dynamic\_inmemory}} and the {\small \texttt{create\_dynamic}} in the CPython source is that {\small \texttt{create\_dynamic\_inmemory}} will ultimately call {\small \texttt{LoadLibraryFromMemory}} from the custom loader instead.

\subsubsection{Python core shared library reference}
\begin{figure}[!htb]
\begin{center}
\includegraphics[scale=0.8]{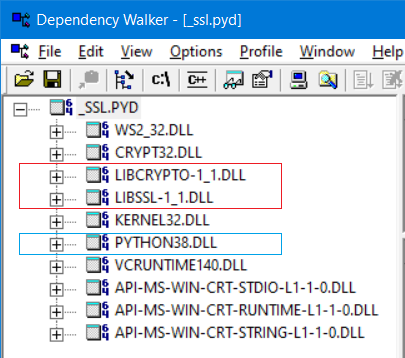}
\end{center}
\caption{\label{fig:python38d} C Extensions may depend on Python core shared library (python38.dll) and other non-system shared libraries. In this example, \_ssl.pyd also depends on OpenSSL's libcrypto and libssl.}
\end{figure}

\begin{figure}[!htb]
\begin{center}
\includegraphics[scale=0.8]{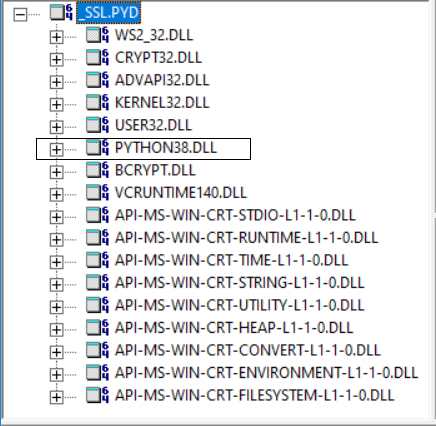}
\end{center}
\caption{\label{fig:python38dnoopenssl} The \_ssl.pyd C Extension statically linked with OpenSSL: The libcrypto and libssl dependencies have been removed. Additional system shared libraries that libcrypto and libssl depended on appear under \_ssl.pyd now.}
\end{figure}

The Python C extensions depend on the Python core shared library (see Figure~\ref{fig:python38d} and Figure~\ref{fig:python38dnoopenssl}), which in our case is already loaded in the process's address-space and may not exist in the path. Therefore, the {\small \texttt{LoadLibraryFromMemory}} implementation in the interpreter that fulfills {\small \texttt{create\_dynamic\_inmemory}} returns the reference of this already loaded shared library when a C extension requests to load this Python core shared library from disk. We also store this reference as \textit{sys.dllhandle}, so that other modules can use it if they need (See Section~\ref{ctypes}).

\subsection{Special case: ctypes}\label{ctypes}
Python's ctypes is a Foreign Function Interface (FFI) library designed to support calling both functions in native shared libraries and also the underlying C functions in the Python interpreter's core shared library via \textit{ctypes.pythonapi}. An example is presented below, where we invoke the Windows system shared library user32.dll's {\small \texttt{MessageBoxA}} to display a modal dialog box.

The \_ctypes package depends on libffi~\cite{libffi} as a shared library on disk. To load \_ctypes into the in-memory interpreter, we statically linked libffi into \_ctypes. This removes the requirement for a libffi shared library to reside on disk. Because ctypes allows invoking C functions in the Python core shared library itself using \textit{ctypes.pythonapi}, we must compare the base address reference of the interpreter against the value we stored in \textit{sys.dllhandle}, and if equal, we use the in-memory loader to resolve the function lookups: This is because a call to \texttt{GetModuleHandle} will not return the base address of the in-memory CPython interpreter's since it was not loaded from disk by the system's \texttt{LoadLibrary} functionality.

\lstset{language=Python, frame=single}
\lstset{label={lst:code_messagebox}}
\begin{lstlisting}[caption=Invoking MessageBoxA from ctypes]
import ctypes
MB_OK = 0
ctypes.windll.user32.MessageBoxA(None,
    b"Hello", b"World!", MB_OK)
\end{lstlisting}

\subsection{Special case: Threading}
To ensure Python code runs from threads created in C, first call {\small \texttt{PyGILState\_Ensure}} to acquire the global interpreter lock (GIL) and store the thread state. Then run Python code. Once done, call {\small \texttt{PyGILState\_Release}} to reset the thread state and release the GIL~\cite{gilstateensure}.

\subsection{Other minor details}
\subsubsection{OpenSSL}
The Windows OpenSSL implementation calls {\small \texttt{GetModuleHandleEx}}, which will not return the expected value since the library was not loaded by the Windows' Loader but our customer loader. Therefore, OpenSSL should be configured with {\small \texttt{-DOPENSSL\_USE\_NODELETE}}, which does not emit the {\small \texttt{GetModuleHandleEx}} code. Any C Extension calling {\small \texttt{GetModuleHandleEx}}, will need to be conditionally compiled to properly account for the case where the extension was not loaded by the Windows' Loader.

\subsubsection{C Runtime  Library}
Similar to the official CPython interpreter, this in-memory CPython interpreter expects Microsoft C Runtime Library v14 (vcruntime140.dll) is installed on the system~\cite{vc140}: A fully-patched version of Windows 10 will already have this runtime DLL.

\subsection {POSIX implementation}
\label{sec:posix}
The POSIX implementation follows the same process; the only distinction is that the custom loader implements {\small \texttt{dlopen\_from\_memory}}. The usual route of in-memory code loading and execution involves allocation of memory pages marked executable via {\small \texttt{mmap}} and {\small \texttt{mprotect}}, however a shortcut is available in most cases. Because POSIX systems present the opportunity to represent many resources as a file, bytes from the shared object can be loaded into a pipe or memory-backed file descriptor and the loading and relocation process can be handled by {\small \texttt{dlopen}} from the C standard library (libc). Once loaded into an executable memory page, the {\small \texttt{dlsym}} provided by the operating system libc yields pointers to the desired functions from the shared object. A modified {dynload\_shlib.c} routes all imports of native code through this mechanism for a complete in-memory experience.

\section{Demonstrations}
\label{sec:evaluation}
The in-memory embedding described in this paper is equivalent in functionality, in every way, to a stock CPython interpreter available for download from Python's official website. Of course, additional functionality can be loaded onto this in-memory embedding using modules or packages stored in ZIP archives, including native code implemented as Python C Extensions. 

One approach threat-actors take is to use an exploit to deploy shellcode to perform actions in the context of an exploited process. Instead of using a 0-day or an \emph{N}-day exploit, we simulated one: We developed an application that instantiates notepad.exe as a child process, writes shellcode into this instantiated notepad.exe's memory, and creates a remote thread to run this shellcode. That is, in this simulation, we are assuming notepad.exe was somehow exploited and a threat-actor successfully wrote shellcode to a location on memory and started a thread to run this shellcode. This shellcode downloads the in-memory Python interpreter and a ZIP archive into memory. The shellcode then loads and initializes the interpreter, adds a ZIP archive in memory with the target module, and calls the module in this archive.

The following demonstrations implement the topics we discuss in this paper, specifically, executing interpreted Python code directly from memory, calling native operating system functions, and loading Python C extensions in-memory and invoking functions in them. The demonstrations present a few well-known security research techniques implemented in Python. The artifacts will be made available as per Section~\ref{sec:availability}. We choose not to offer any new malware or reimplement existing malware into Python to run on top of this embedding: We are extremely wary about repurcussions of publishing malware along with this paper that security products may or may not defend against. 

\subsection {A simple demonstration covering all topics}
In this demonstration, the module invokes the code in Listing~\ref{lst:code_messagebox} and produces a ``Hello, world!'' Message Box by calling an operating system C function. This demonstrates all the concepts discussed. Specifically, in this demonstration, we showed the CPython embedding executing interpreted Python code, loading a Python C Extension (ctypes), and then invoking an operating system C function \texttt{MessageBoxA} exported by the Windows shared library user32.dll.

\subsection {Enumerate all processes on system}
In this demonstration, we enumerate all processes by making a call to the partially documented Windows operating system function \texttt{ntdll!NtQuerySystemInformation}. This is a common technique malware use for reconnaissance: Enumerate processes and determine if any may indicate the process is emulated either in a virtual machine or if the system has a security product such as an antivirus is installed. This example also demonstrated all the concepts discussed: We loaded Python C Extension ctypes, and then invoked the operating system C function exported by Windows shared library ntdll.dll.

\subsection {Download a file from the Internet}
In this demonstration, using the urllib.request package, we download a file from a website. This also demonstrates all the concepts discussed. The urllib.request packages utilizes the socket and ssl packages. The socket and ssl packages import \_socket.pyd and \_ssl.pyd Python C Extensions, respectively. The extension \_ssl.pyd statically links with OpenSSL. Sophisticated malware usually deploys in stages: The first stage performs reconnaissance, e.g., process enumeration. If the malware determines it is in the desired environment, it then reaches out to the Internet and downloads additional stages. In this demonstration we used urllib.request package to download a file from the Internet.

\subsection {A capability to use the system BCrypt library}
In this demonstration, we call cryptographic functions in the Windows operating system's BCrypt library from Python. This is something a late stage Malware may use, for example, to decrypt users' website credentials stored in Chrome's sqlite databases.

We offered a few samples demonstrating techniques security researchers employ. Given that this an in-memory embedding of a stock CPython, any code that would work in the stock CPython will also work in this embedding.

\section{Future Work}
\label{sec:future}
Some third-party Python packages not shipped with Python itself, such as PyCryptodome, require additional work to load into the in-memory interpreter. Specifically, files such as \_raw\_cbc.pyd, \_raw\_cfb.pyd, etc., are not proper Python C Extensions: These cannot initialize as C extensions as defined per {\small \texttt{FAKE\_INIT}}~\cite{pycryptodome}. PyCryptodome uses the \textit{cffi} package to load these, and \textit{cffi} expects the Python C extensions to reside on disk. The \textit{cffi} package does not come standard with CPython. We may fork this \textit{cffi} package, such that the requirement that these pseudo extension files need to be on disk is removed. The number of packages dependent on \textit{cffi} can be identified in \url{https://libraries.io/pypi/cffi/dependents}. Essentially, any third-party library that offers a C Foreign Function Interface (CFFI) and wants to support in-memory dynamic loading will need to implement a functionality similar to~\cite{fancycode}. We demonstrated our approach for CPython 3.8. This approach has also been validated to work in CPython 3.7 and CPython 3.9. We intend to support future versions of CPython, as the changes to apply are the same. We believe our in-memory CPython embedding has utility beyond offensive computer security research given it's simplicity in containing the entire CPython interpreter and default libraries in a single shared library.


\section{Conclusion}
\label{conclusion}
Red Teams want to prototype malware from different threat-actors for their campaigns, which is difficult to accomplish rapidly when writing in C or assembly. Python is a popular interpreted high-level language, and would allow for such rapid prototyping. In this paper, we describe an in-memory embedding of CPython, which can be used for this type of prototyping, and can accomplish any specific task needing C or assembly via Python C extensions. We mentioned examples of malware written in Python and existing embeddings of the CPython interpreter that required the interpreter's files to reside on disk. We then detailed changes that allowed the CPython interpreter to run entirely in memory without ``touching'' the disk, and also loading Python packages and modules into it directly from memory. Finally, we covered handling special cases such as the \textit{ctypes} package and threading. We believe our approach to embedding is simple and therefore, has utility beyond offensive computer security.



\section*{Availability}
\label{sec:availability}
The artifacts submitted for evaluation are provided in Appendix A. The complete CPython source code with our modifications is available on GitHub under \url{https://www.github.com/scythe-io/in-memory-cpython}.

\appendix[Artifact Readme]
\label{appendix-artifact-readme}
This appendix describes the steps necessary to download a harness-exe that performs the demonstrations listed in the Demonstration section. Depending on the specific demonstration, the harness either creates a notepad.exe child process or uses the current console process. The harness-exe process allocates some Write+Execute memory, copies some shellcode and a harness DLL and starts a thread to execute the shellcode. The shellcode loads the harness DLL, and this downloads a CPython DLL constructed as described in the paper, and a demonstration chosen by the user in the steps of harness-exe. The specific steps involving the shellcode and the harness DLL emulate what would happen after an 0-day or N-day exploit without using an actual 0-day or N-day or compromising a system. Because this harness-exe is downloaded from the Internet, it will have a ``Mark-of-the-Web," and Windows Defender will quarantine when you run it. Therefore, please follow the instructions in this document to Add an Exclusion for the harness-exe in Windows Defender.

\subsection{Artifact Checklist}
\noindent Binary: harness-exe.exe \\
Run-time environment: verified on Windows 10 x64 \\
Hardware: A commodity PC is sufficient \\
Disk space required approximately: 100GB \\
Time needed to prepare workflow: 60 minutes \\
Time needed to complete experiments: 30 minutes \\
Publicly available: \url{https://doi.org/10.5281/zenodo.4638251} and also under \url{https://github.com/farfella/woot2021/} \\
Code licenses: zlib 

\subsection{Description}
The harness-exe has been verified to work on Microsoft’s Windows 10 x64 Edge VM for VMWare. These instructions assume using Microsoft’s Edge VM. Step 6 of the Installation section covers how to retrieve and execute harness-exe.

\begin{enumerate}
\item Download \emph{VMWare Workstation Player} (approximately 215MB) and install on your computer (free for non-commercial use): \url{https://www.vmware.com/products/workstation-player/workstation-player-evaluation.html}
\item Download \emph{MSEdge on Win 10 (x64) Stable 1809} (approximately 6.7GB) from: \url{https://developer.microsoft.com/en-us/microsoft-edge/tools/vms/}. Choose VMWare for the VM platform.
\item Extract the downloaded MSEdge.Win10.VMware.zip and open the \texttt{MSEdge-Win10-VMware.ovf} inside the extracted folder MSEdge-Win10-VMware.
\item This will start VMWare Workstation Player importing the Edge Virtual Machine (VM).
\item Log into this Edge VM using:
\begin{enumerate}
\item User: \texttt{IEUser}
\item Password: \texttt{Passw0rd!}
\end{enumerate}
\item In this VM, create a folder named \texttt{woot2021} on the C drive.
\item Start an \emph{Administrative Command Prompt}.
\item From the Start menu, type \texttt{cmd}, and when Command Prompt appears, choose \emph{Run as administrator} on the right as shown in the figure below and approve the consent dialog box.
\item Type \emph{powershell} and hit Enter to start a Powershell prompt:
\item Add \texttt{C:\textbackslash woot2021} to the Windows Defender Exclusions, turn off automatic sample submission, and download and install Visual Studio redistributable (approximately 1MB). The installation of the redistributable is a requirement for CPython itself. To accomplish these tasks, copy-and-paste the following commands into the Powershell prompt (Step 5) and hit Enter:
\item Install the redistributable as shown below.
\item Download the \texttt{harness-exe} zip also using Powershell and open the folder. Type the following and hit Enter.
\begin{verbatim}
Invoke-WebRequest
-Uri "https://github.com/farfella/
      woot2021/raw/master/
      harness/harness-exe.zip"
-OutFile "C:\woot2021\harness.zip"
explorer .
\end{verbatim}
\item Extract the archive into the C:\textbackslash woot2021\textbackslash harness folder. The password to extract is: \texttt{w00t-2021-w00t}
\end{enumerate}

\subsection{Experiment workflow}
In the same command prompt as Installation steps, change directory to c:\textbackslash woot2021\textbackslash harness. Then run .\textbackslash harness-exe.exe and hit Enter.

\begin{verbatim}
  CD C:\woot2021\harness
  .\harness-exe.exe
\end{verbatim}

When you run harness-exe.exe, some instructions are presented. After this, the reviewer may choose the default URL in which the in-memory CPython DLL resides (e.g., \url{https://github.com/farfella/woot2021/tree/master/in-memory-embedding-cpython}) or download from this GitHub repository and offer the DLL from the reviewer’s own hosted server, for example.

Following this step, the reviewer is prompted to choose one of five demonstration options. Options one through four are the examples described in the Demonstrations section in the paper:
\begin{enumerate}
\item A demonstration covering all topics (spawns child notepad.exe and shows a dialog box)
\item A demonstration that prints all the process names running on the system
\item A demonstration that downloads a file from the Internet
\item A demonstration calling BCrypt to encrypt and decrypt a string
\end{enumerate}

You will need to run harness-exe.exe for each demonstration.

\subsection{Evaluation and expected results}
For option 1, a child notepad.exe will be spawned, in which Write+Execute memory will be allocated, and shellcode and harness DLL will be written. Then a thread will be spawned in this child notepad.exe to execute this shellcode. The shellcode will load the harness DLL. The harness DLL will retrieve the CPython DLL constructed as described in the paper and also the zip file corresponding to demonstration 1 from either the default URLs or reviewer-provided URLs.

For option 2, 3, and 4, in the current process (i.e., process of harness-exe.exe) the write+execute memory will be allocated and the shellcode and harness DLL will be written. This is because these examples use print() to write to standard output, and the current process is a console application that can support this functionality. The CPython DLL is instantiated with verbosity turned on (i.e., equivalent to running python\~-v). As such, you will see debug messages on the screen to give you more insight.

\subsection{Experiment Customization}
The fifth demonstration option allows you to offer a zip file with a module named magic inside (i.e., magic.py or magic.pyc) from their own hosted URL. The harness DLL is constructed to only \texttt{import magic}. This allows you to test with some custom Python code instead of the four demonstration options. Note however, that only the standard Python packages are embedded in this CPython DLL. The demonstrations are available under: \url{https://github.com/farfella/woot2021/}.

\bibliographystyle{plain}
\bibliography{library}

\end{document}